# UGC 7639: a Dwarf Galaxy in the Canes Venatici I Cloud

L.M. Buson,[1] D. Bettoni,[1] P. Mazzei,[1] and G. Galletta[24]

[1] INAF Osservatorio Astronomico di Padova, vicolo dell'Osservatorio 5, I 35122 Padova, Italy
e-mail lucio.buson@oapd.inaf.it, daniela.bettoni@oapd.inaf.it, paola.mazzei@oapd.inaf.it

[2] Dipartimento di Fisica e Astronomia, vicolo dell'Osservatorio 2, I-35122 Padova, Italy
e-mail: giuseppe.galletta@unipd.it

**ABSTRACT**

.

We want to get insight into the nature, *i.e.* the formation mechanism and the evolution, of UGC 7639, a dwarf galaxy in the Canes Venatici I Cloud (CVnIC).
We used archival GALEX (FUV and NUV) and SDSS images, as well as Hyperleda and NED databases, to constrain its global properties. GALEX FUV/NUV images show that UGC 7639 inner regions are composed mostly by young stellar populations. In addition, we used smoothed particle hydrodynamics (SPH) simulations with chemo-photometric implementation to account for its formation and evolution. UGC 7639 is an example of blue dwarf galaxy whose global properties are well matched by our multi-wavelength and multi-technique approach, that is also a suitable approach to highlight the evolution of these galaxies as a class.
We found that the global properties of UGC 7639, namely its total absolute B-band magnitude, its whole spectral energy distribution (SED), and its morphology are well-matched by an encounter with a system four times more massive than our target. Moreover, the current star formation rate (SFR) of the simulated dwarf, ≈0.03 $M_\odot$ yr$^{-1}$, is in good agreement with our UV-based estimate**.**
For UGC 7639, we estimated a galaxy age of 8.6 Gyr. Following our simulation, the ongoing star formation will extinguish within 1.6 Gyr, thus leaving a red dwarf galaxy.

**Key words.** galaxies: ultraviolet — galaxies: dwarf — galaxies: individual: UGC 7639

## 1. Introduction

Dwarf galaxies (DGs) are the predominant class in the total galaxy population of the universe. They are, by definition, galaxies of low mass and size, and could be considered as the smallest baryonic counterparts of the dark matter (DM) constituent blocks in the Universe (*Gilmore et al. 2007; Governato et al. 2010; Adams et al. 2014*). Despite their abundance, we can study in detail only the objects that we see in the local universe, due to their intrinsic faintness. In particular, in the nearby Local Volume (LV), the most detailed information available about DGs comes from the LV population (*cf. Tolstoy et al. 2009; Crnojevic et al. 2011*). However, many recent studies have started to look beyond the LV in order to derive the properties of DGs belonging to several nearby groups. These studies include Cen A, NGC 1407, Coma I, Leo, NGC 1023, M81, Sculptor and Canes Venatici Cloud itself (*e.g.: Trentham et al. 2006; Sharina et al. 2008; Weisz et al. 2008; Bouchard et al. 2009; Crnojevic et al. 2011*).

DGs possess a large variety of morphological types (*Grebel 2001*), the most frequent in galaxy groups being dwarf elliptical (dE), dwarf spheroidal (dSph), and dwarf irregular (dIrr) galaxies (*Grebel, 2001; Lora et al. 2015*).

Among them, dIrrs are galaxies showing recent star formation, exhibiting substantial gas fractions, and a broad range of star formation rates (SFR). The majority of dIrrs belongs to groups rather than being found in the field (*Saviane et al. 2008*).

The evolutionary properties of dwarf galaxies in nearby groups are still not well understood. Their study can help to understand the role of the environment and interactions on galaxy evolution. Moreover, these galaxies can reasonably be considered local analogs of the furthest (*i.e.* youngest) actively star-forming systems.

Our goal is to investigate the evolution of a specific dIrr in a group, UGC 7639. This is a dwarf galaxy belonging to the relatively large number of nearby, low surface brightness galaxies investigated in the context of the recent imaging surveys of the LV neighborhood. The distance of UGC 7639 has been reliably estimated by means of the surface brightness fluctuation method by Rekola *et al.* (*Rekola et al., 2005*). They estimated D=7.1±0.6 Mpc, in excellent agreement with the value (7.14±0.5 Mpc) given in the updated nearby galaxy catalogue of Karachentsev *et al.* (*Karachentsev et al.,* 2013). UGC 7639 is a member of the Canes Venatici Cloud (*Tully, 1988*). The so-called Canes Venatici Cloud consists of two complexes aligned along the same line of sight, including mainly late-type DGs. More specifically, it is formed by CVnI Cloud and CVnII Cloud, at $V_r$=300 km s$^{-1}$, *i.e.*, at an average distance of 7.68±0.9 Mpc, and $V_r$=560 km s$^{-1}$, *i.e.*, at an average distance of 17.2±1.3 Mpc, respectively (cf. *Makarov et al. 2013*). As a consequence, UGC 7639, with a measured redshift of 382 km s$^{-1}$ (*Rekola et al. 2005*) likely belongs to the CVnI Cloud.

UGC 7639 caught our attention because its central regions, as recorded by the GALEX satellite, appear dominated by FUV emission (Fig. 1, left panel). This object likely hosts an amount of young stars typical of an ongoing starburst. This is also a general property of DGs of different morphological types (*Grebel, 2001; Morrissey et al. 2007; e.g. Koleva et al. 2014*).

In the past, several different morphological classifications have been proposed for this object. deVaucouleurs *et al.* (1991) (RC3) and later *Rekola et al.* (2005) proposed a transition class dE/Im. Conversely, a dS0/BCD (Blue Compact Dwarf) classification has been adopted by Bremnes et al (2000) and Parodi & Binggeli (2003) considering its outer diffuse elliptical halo and inner multiple star-forming regions and star clusters. Finally, *Karachentsev et al.* (2013) favor a true Im classification. Indeed, the galaxy shows a pattern of inner star forming regions easily seen in the SDSS composite image (Fig. 1, right panel), although the Hα imaging by *Kaisin and Karachentsev* (2008) barely supports this last classification.

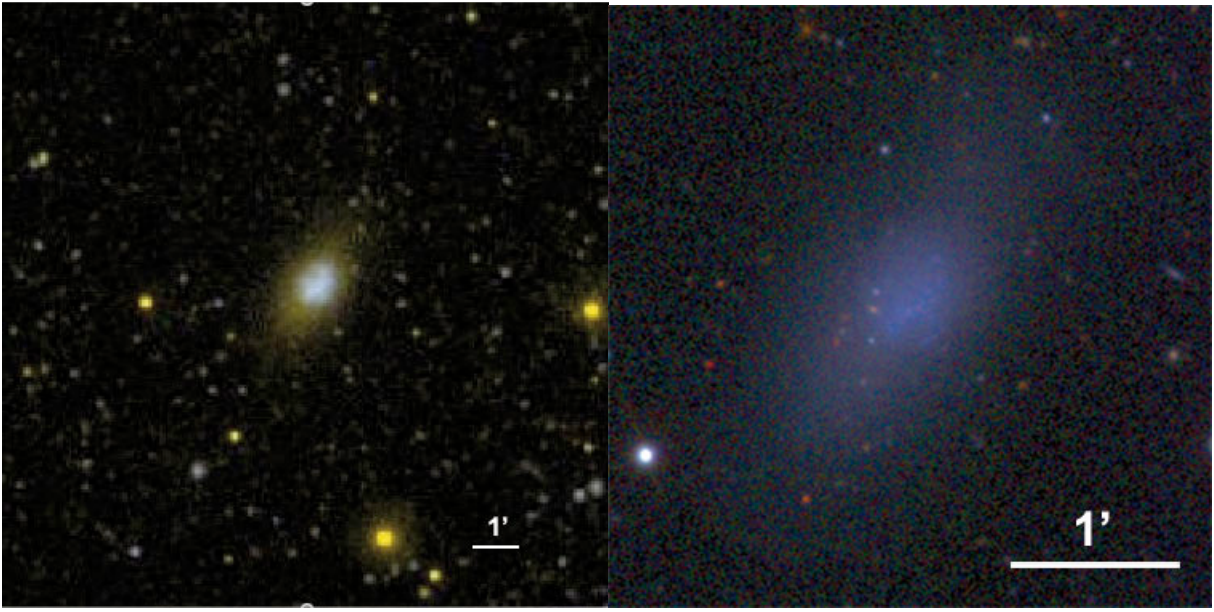

Fig. 1. Left: Composite 13' × 13' GALEX image of UGC 7639. Blue and yellow correspond to FUV and NUV. Right: Composite 3'.6 × 3'.6 SDSS (gri) image of UGC 7639. Note the outstanding, innermost blue patches, which are also weak Hα emitters.

In this paper, we want to investigate the formation mechanism and the evolution of this dwarf galaxy, as an example of DG evolution.

We used GALEX (FUV and NUV) and SDSS g and r-band images as well as Hyperleda and NED data to constrain its global properties. In addition, smoothed particle hydrodynamics (SPH) simulations with chemo-photometric implementation gave us insight into its formation mechanism, together with the past and future evolution of this galaxy. These SPH simulations allowed us to derive dynamical and morphological information, concurrently with the spectral energy distribution (SED) extended from far-UV to 1 mm, at each evolutionary time (snapshot, hereafter). Finally, they allowed us to trace the evolutionary path of the selected galaxy in the optical and UV color-magnitude diagram, (NUV-r) vs $M_r$.

The present work is organized as follows: in Section 2, archive multi-wavelength data are analyzed and physical properties of our target are derived. The UV-optical color–magnitude diagram (CMD) of all 22 members of the Canes Venatici I Cloud, to which our target belongs, is also examined to characterize the surrounding environment. Section 3 focuses on the description of the adopted modeling, and on our results. Finally, conclusions are summarized and discussed in Section 4.

The selected dwarf, with a reliable distance measured, a large wavelength coverage of the observed

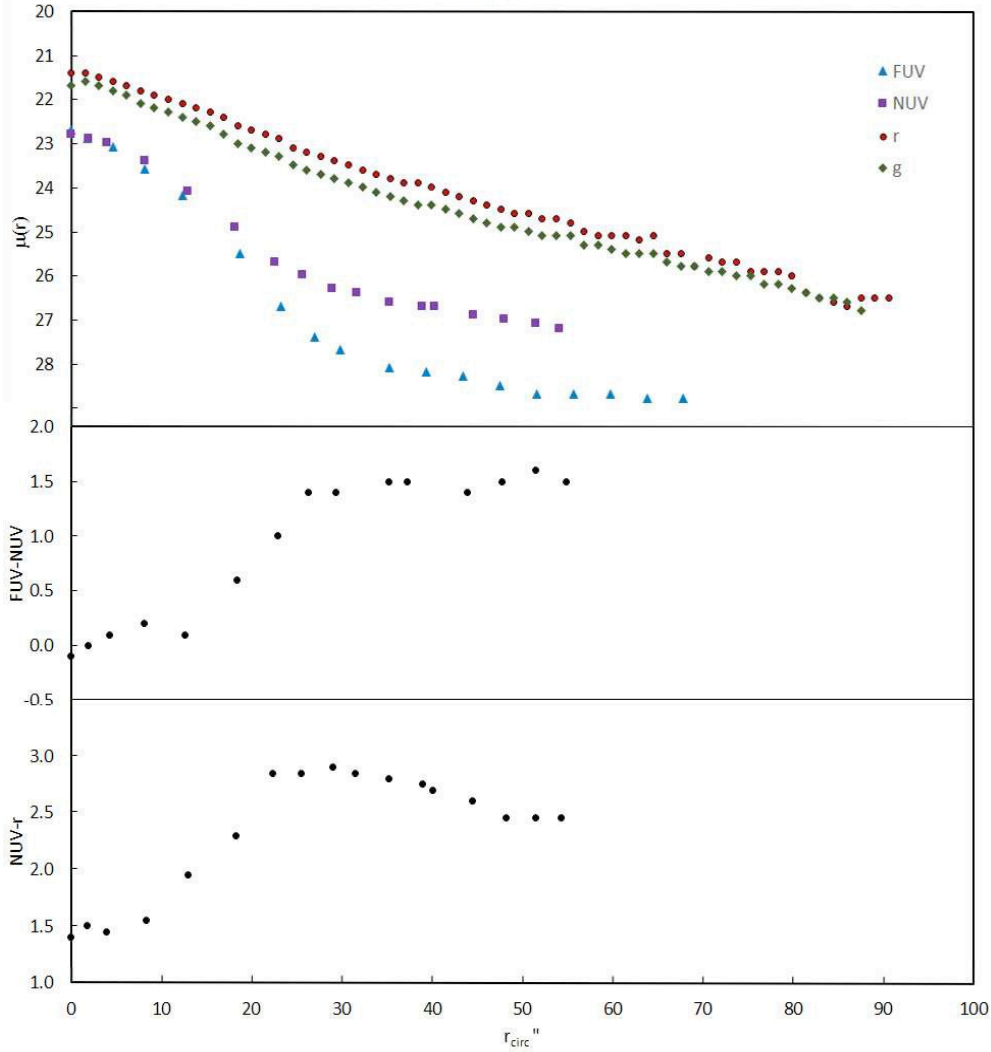

**Fig. 2.** Top panel: GALEX NUV and FUV, optical r and g-band surface brightness profiles of UGC 7639 (magenta squares, cyan triangles, red and green circles, respectively) *vs.* circularized radius. Middle panel: (FUV-NUV) color profile. Bottom panel: UV/optical (NUV-r) color profile. Since BCDs experience mostly off-centered star formation (cf. *Koleva et al. 2014*), these profiles could be affected by the difficult identification of the main stellar body centre.

SED (Sect. 3), and other properties discussed in the following section (Sect. 2), provides an useful example of this class of objects, helping us to constrain our simulations and giving insights into the formation mechanism and the evolution of this dwarf as an example of DG evolution.

## 2. Physical parameters.

### 2.1. UV/Optical Photometry

To study both the UV and the optical morphology, we extracted two UGC 7639 images (3249s and 5883s for FUV and NUV cameras, respectively) from the GALEX archive (they belong to tile G11_47072_UGC07639, obtained as part of the program GI1 047 P.I. R. Kennicutt) together with r and g-band images belonging to the SDSS-DR7 archive (*Abazajian et al.*

2009).

From these images we performed the surface photometry by means of the IRAF STSDAS ELLIPSE routine for NUV, FUV, g and r bands. ELLIPSE computes a Fourier expansion for nested isophotes (*Jedrzejewski 1987*), producing the surface photometric profiles. The resulting UV photometric profiles and corresponding (FUV-NUV), (NUV-r) color profiles *vs.* circularized radius are shown in Fig. 2. The circularized radius has been derived from the relation $r_c = a_e * (1-\varepsilon)^{1/2}$, were $a_e$ is the radius measured along the ellipse major axis and $\varepsilon$ is the measured galaxy ellipticity. The average colors (derived inside the full extent of the galaxy in the GALEX images: ~60") are: (FUV-NUV)=0.40±0.07 and (NUV-r)=2.58±0.18 which are typical of late type DGs (*Wyder et al. 2007*). However in its innermost region (10 arcsecs i.e. ≈ 0.4 kpc), the galaxy shows much bluer colors, *i.e.* (FUV-NUV)=0.06±0.11 and (NUV-r)=1.47±0.06 (Fig.. 2, bottom panels) a characteristic of dIrrs (*Hunter and Elmegreen 2006*).

Our resulting g and r magnitudes, effective radii, and average surface brightnesses are reported in Table 1. Taking into account the classical Fukugita *et al.* (1996) conversion formulae, we compared our integrated g and r AB magnitudes with those of Bremnes *et al.* (2000) in the Johnson B and R bands, *i.e.* $m_B$=13.94±0.21 and $m_R$=12.93±0.20 mag, finding a good agreement.

We found that the best agreement with the observed g and r profiles is obtained with a pure disk component. The FUV and NUV profiles show a light excess in the central region, due to UV spots (see Figure 1), that cannot be fitted with a smooth component. For this reason we give here only the total magnitudes, NUV=15.7±0.04 and FUV=16.1±0.04. These integrated magnitudes, in good agreement with the values NUV=15.71±0.03 and FUV=16.13±0.05 measured by Lee *et al.* (2011), will be used to constrain our simulations (Sect. 3.1.1).

Table 1 Measured optical photometric parameters

| Band | $r_e$ | $\mu_e$ | $m_{tot}$ | $M_{tot}$ |
|---|---|---|---|---|
| | arcsec | mag arcsec$^{-2}$ | mag | mag |
| r[AB] | 30 | 22.9 | 13.12 ±0.08 | -16.14 |
| g[AB] | 31 | 23.6 | 13.51±0.10 | -15.75 |

2.2. The environment

As already pointed out, UGC 7639 belongs to the CVnI Cloud that includes 22 galaxies (*Tully, 1988*). To characterize UGC 7639 inside this Cloud we obtained, for each galaxy, the available SDSS r-band magnitudes from the NASA-Sloan Atlas (*Blanton et al. 2011*), and the NUV and FUV data from Lee *et al.* (2011). Extinction corrections have been extracted, for each galaxy, from the same NASA-Sloan Atlas (*Blanton et al. 2011*).

For UGC 7639 we used the updated distance modulus by Rekola *et al.* (2005), while for the other galaxies we used the distances given by Tully (1988). Using these data, we derive the color-magnitude diagram (CMD) (NUV-r) vs. $M_r$ (*Martin et al. 2007; Wyder et al. 2007; Fang et al. 2012*), showed in Fig. 3. All but four galaxies follow the blue sequence (cyan line), and no galaxy lies on the red sequence, typical of evolved and/or reddened systems. NGC 4096, NGC 4248, NGC 4460, and NGC 4707 are in the intermediate region, the so-called green valley. Following the classification of Tolstoy *et al.* (2009), there are eight DGs (i.e. $M_B > -16.0$ mag) in this Cloud, accounting for ~40% of its member galaxies, and looking at Fig. 3, UGC 7639 is among the faintest members of this group.

The UV luminosity is a tracer of the present-day SFR, that can be derived following Kennicutt (1998), from the relation:

$$SFR_{FUV} \ (M_\odot yr^{-1}) = 1.4 \times 10^{-28} L_{FUV} \ (ergs \ s^{-1} \ Hz^{-1})$$

Using the $L_{FUV}$ from FUV magnitudes that we measured (Sect. 2.1), and our adopted distance, the SFR turns out to be $\sim 10^{-2} M_\odot \ yr^{-1}$ for our dwarf, in agreement with the value found by Karachentsev and Kaisina (2013). Such relation accounts for a Salpeter's IMF with lower and upper mass limits 0.1 $M_\odot$ and 100 $M_\odot$, respectively. The value of the SFR here derived will be used to further constrain our simulation (Sect. 3.1.1).

Since total UV magnitudes are available for each member of the Cloud (*Lee et al. 2011*), using the previous formula we highlight with different symbols in Figure 3 the different levels of the present-day SFR of the galaxies in this Cloud. Only few members harbor a SFR above 0.1 $M_\odot$/yr (squares). NGC 4258, the brightest galaxy in such figure, reaches the highest value, namely 0.6 $M_\odot$/yr. The Tully (1988) catalog also provides the amount of cold gas (HI) in 19 out of 22 galaxies in the Cloud. We complete our sample by adding data of NGC 3985 and NGC 4460 from Springob *et al.* (2005), and of NGC 4485 from the RC3 catalog, respectively. UGC 7639 contains the lowest amount of neutral gas ($\sim 4 \times 10^7 M_\odot$) in the Cloud, while NGC 3985 contains the highest amount, namely $\sim 10^{11} M_\odot$. The amount of cold gas in our DG constrained further our simulations (Sect. 3.1.1).

## 3. Modeling

The novelty of our approach is that we explore the evolutionary scenario of our selected galaxy using a large set of SPH simulations of galaxy encounters and/or mergers, including a chemo-photometric code based on evolutionary population synthesis **(EPS)** models. The general prescriptions of SPH simulations and the grid of impact parameter explored, reported in several previous papers (*e.g. Mazzei et al 2014a,b*), are summarized below.

Our SPH simulations of galaxy formation and evolution start from the same initial conditions described in Mazzei & Curir (2003, MC03 hereafter) and Mazzei (2003, and references therein), *i.e.* collapsing triaxial systems composed of DM and gas with density distribution $\rho \propto r^{-1}$, in different proportions and different total masses. All of the simulated systems have the same initial virial ratio (0.1), as well as the same average density and spin parameter. In more detail, each system is built up with a spin parameter, $\lambda$, given by $|J||E^{0.5}|/(GM^{0.5})$, where E is the total energy, J is the total angular momentum, and G is the gravitational constant; $\lambda$ is equal to 0.06 and aligned with the shortest principal axis of the DM halo. The triaxiality ratio of the DM halo, $\tau = (a^2 - b^2)/(a^2 - c^2)$, is 0.84, where a > b > c (Warren et al. 1992).

We then produced a large set of galaxy encounters involving systems with a range of mass ratios from 1:1 to 1:10. In order to exploit the vast range of orbital parameters, we carried out different simulations for each pair of interacting systems, varying the orbital initial conditions in order to have, for an ideal Keplerian orbit of two mass points, the first pericentre separation (*p*) ranging from the initial length of the major axis of the DM triaxial halo of the primary system to 1/10 of the same (major) axis. For each of these separations, we changed the eccentricity in order to have hyperbolic orbits of different energies. For the most part, we studied direct encounters, where the spins of the systems are equal (MC03), generally parallel to each other, and perpendicular to the orbital plane. However, we also analysed some cases with misaligned spins in order to enhance the effects of the system initial rotation on the results. Moreover, for a given set of encounters with the same orbital parameters, we also examined the role of increasing initial gas fractions.

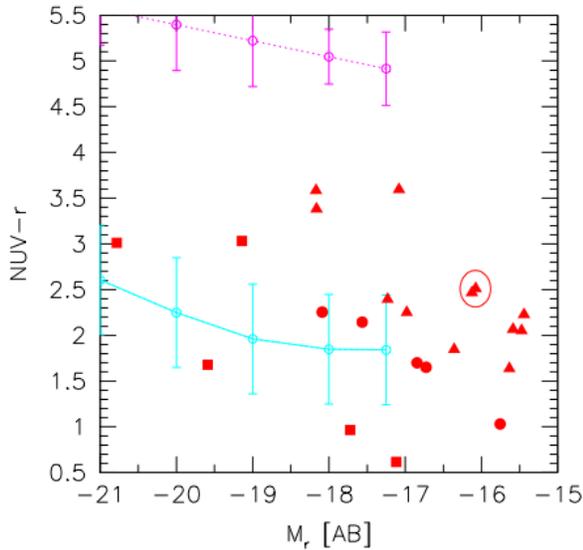

**Fig. 3.** The CMD (NUV-r) vs. $M_r$ of the CVnI Cloud members. The Wyder *et al.* (2007) fits of the blue and red sequences (cyan/magenta lines) are plotted, including their error bars. The oval indicating the position of UGC 7639 also includes UGC 7408, a galaxy slightly bluer and brighter than our dwarf. Different symbols correspond to different SFRs: squares are for SFR≥0.1 $M_\odot$/yr, triangles for SFR≤0.05 $M_\odot$/yr and circles for SFRs in between.

All of the simulations include the self-gravity of gas, stars and DM, radiative cooling, hydro-dynamical pressure, shock heating, viscosity, star formation, feedback from evolving stars and type II supernovae, and chemical enrichment. Simulations provide the SED at each evolutionary time, *i.e.*, at each snapshot. The time step between individual snapshots is 0.037 or 0.151 Gyr, as for the simulation which best fits the global properties of UGC 7639 (see below). The SED we derived accounts for chemical evolution, internal extinction and re-emission by dust in a self-consistent way, and extends at least over four orders of magnitude in wavelength, *i.e.* from 0.06 to 1000 µm. The SED is based on EPS models exploiting isochrones, and this method has been fully described in several previous papers (*Mazzei et al. 1992, 1994, 1995*). Mazzei *et al.* (1992) in particular was the first paper where the SED is extended over four orders of magnitude in wavelength (*Lonsadale, 1993*). SPH simulations with chemo-photometric implementation were presented for the first time in Curir & Mazzei (1999). In these papers, authors improved chemo-photometric predictions, accounting for six stellar populations, from star metallicity 0.0004 to 0.05; a stellar population entails, by definition, all star clusters (*i.e.*, particles) born with the same chemical compositions.

Each simulation self-consistently provides morphological, dynamic and photometric evolution. The initial mass function (IMF) is of Salpeter type (*Salpeter 1955*) with upper and lower mass limit 100 and 0.01 $M_\odot$ respectively, as in *Curir & Mazzei* (1999) and MC03.

All the model parameters have been tuned in these previous papers that analyzed the evolution of isolated collapsing triaxial halos initially composed of DM and gas. In these papers, the role of the initial spin of the halos, their total mass and gas fraction, as well as different IMFs, particle

Table 2. **Input parameters of SPH simulation of UGC 7639**

| $N_{part}(t=0)$ | a [kpc] | p/a | $r_1$ [kpc] | $r_2$ [kpc] | $v_1$ [km/s] | $v_2$ [km/s] | $M_1$ $10^{10} M_\odot$ | $M_2$ $10^{10} M_\odot$ |
|---|---|---|---|---|---|---|---|---|
| $10^5$ | 597 | 1/4 | 117 | 549 | 12 | 48 | 40 | 10 |

**Notes.** The columns are as follows: Col. (1) total number of particles; (2) length of the semi-major axis of the primary halo; (3) pericentric separation of the halos in units of semi-major axis in col. (2); (4) and (5) distances of the halo centers of mass from the centre of mass of the global system; (6) and (7) velocity moduli of the halo centers in the same frame; (8) and (9) total masses of each system.

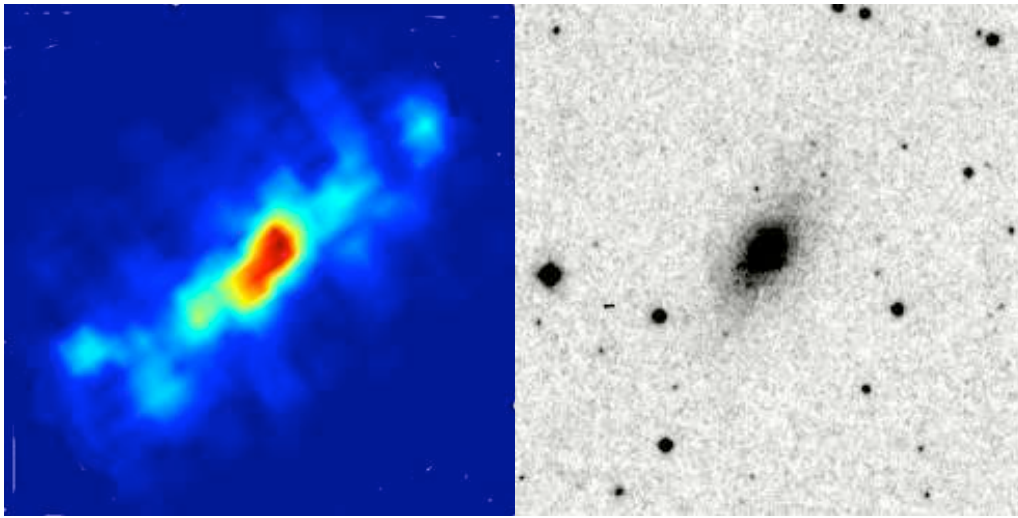

**Fig. 4.** Morphology of UGC 7639. Left: R-band luminosity density map from our simulation at the best-fitting snapshot (see text), on the same scale as the observed one. Right: 7' × 7' R-band image (dss1). Left map is normalized to its total flux.

resolutions, SF efficiencies, and values of the feedback parameter, were all examined. The integrated properties of simulated galaxies, stopped at 15 Gyr, *i.e.*, their colors, absolute magnitudes, metallicities and mass to light ratios, had been successfully compared with those of local galaxies (*Curir & Mazzei 1999*, Figure 17; *Mazzei 2003, 2004*, Figure 8). In particular, a slightly higher SFR, compared with the other possibilities examined in MC03, arises from our IMF choice (see MC03: Fig. 1); this allows for the lowest feedback strength (63% less than in the same simulation with lower mass limit 0.1 $M_\odot$), and for the expected rotational support when disk galaxies are formed (MC03).

As pointed out by Kroupa (2012), this slope is almost the same as the universal mass function that links the IMF of galaxies and stars to those of brown dwarfs, planets, and small bodies (meteoroids, asteroids; *Binggeli & Hascher 2007*).

We point out that the SFR, which drives the evolution of the global properties of the simulated galaxies, converges when the initial particle number is above $10^4$ (see MC03 for a discussion, their Figure 1; *Christensen et al. 2010, 2012*).

From our grid of SPH simulations, we singled out those that simultaneously account (*i.e.* at the same snapshot) at least for the following three observational constraints, which correspond to the global properties of UGC 7639:

1) total B-band absolute magnitude within the range allowed from observations (Sect. 2.1);
2) best-fit of the integrated observed SED;
3) morphology matching the observed one in the same bands and with the same spatial scale (arcsec/kpc).

We direct the interested reader to the papers by Mazzei *et al.* (2014a) - where our approach has been applied to early-type galaxies of two groups, USGC 376 and LGG 225 - and Mazzei *et al.* (2014b), where our approach has been exploited to match not only photometric, but also structural (*e.g.* disk versus bulge) and kinematical (gas versus stars) properties of two S0 galaxies, namely NGC 3626 and NGC 1533, to predict their evolution.

The initial conditions of the simulation selected correspond to an encounter: these conditions are shown in Table 2. The initial gas fraction is 0.11 in both the systems. This value is very similar to the value of 0.13 found by Gonzalez *et al.* (2013) by analyzing a large sample of clusters with different total masses. The initial gas mass resolution is $10^6 M_\odot$, while that of DM particles is nine times larger. The gravitational softening is 1, 0.5, and 0.05 kpc for DM, gas and star particles, respectively. The final number of particles at least doubles the initial number in Table 2.

The results presented in the next sections are the predictions of the simulation which best reproduces all the previous conditions (points 1-3) at the same snapshot. This snapshot sets the age of the simulation and the age of the galaxy. This means that the onset of the star formation is delayed compared to the beginning of the simulation, so that the age of the simulation is higher than the galaxy age.

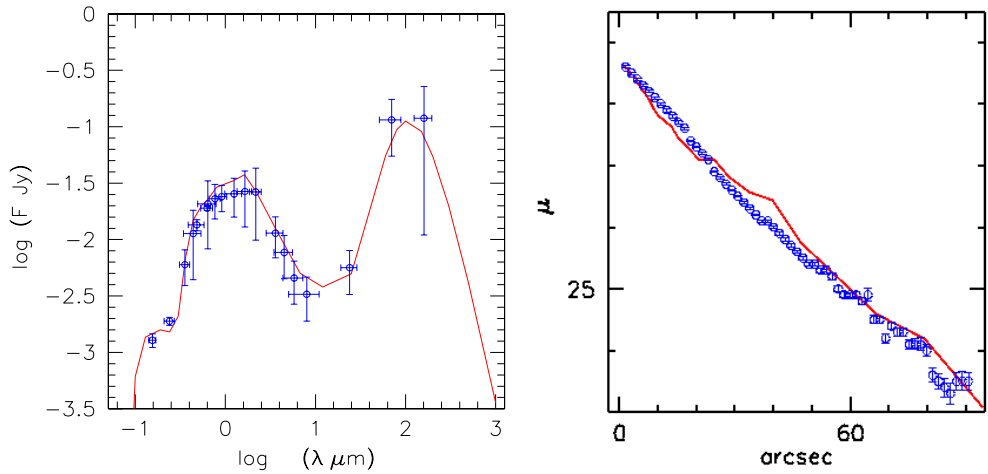

**Fig. 5. Left**: The SED of UGC 7639. Blue filled circles corresponds to UV/Optical/IR/FIR data, in particular FUV and NUV total fluxes from Table 1, u, g, r, i, z, fluxes from NASA-Sloan Atlas (*Blanton et al. 2011*), B and R from Bremnes et a. (2000), IR and FIR from Dale *et al.* (2009). Error bars account for bandwidth and $3\sigma$ flux uncertainties. The solid (red) line shows the prediction of our model (see text). Right: The r-band luminosity profile of our simulated image in Fig.4 is compared with that measured in the same band and shown in Fig. 2. Error bars <1% do not appear in the figure.

3.1. Results

3.1.1. Comparison with observations

The global properties of UGC 7639, *i.e.* its total B-band absolute magnitude, integrated SED and morphology, are well matched by a simulation corresponding to an encounter, not a merger, of two systems with mass ratio 4:1 and total mass $5 \times 10^{11} M_\odot$ (Table 2 ); UGC 7639 corresponds to the less massive system.

The age of the galaxy at the snapshot which best fits the global properties of UGC 7639 is 8.6 Gyr,

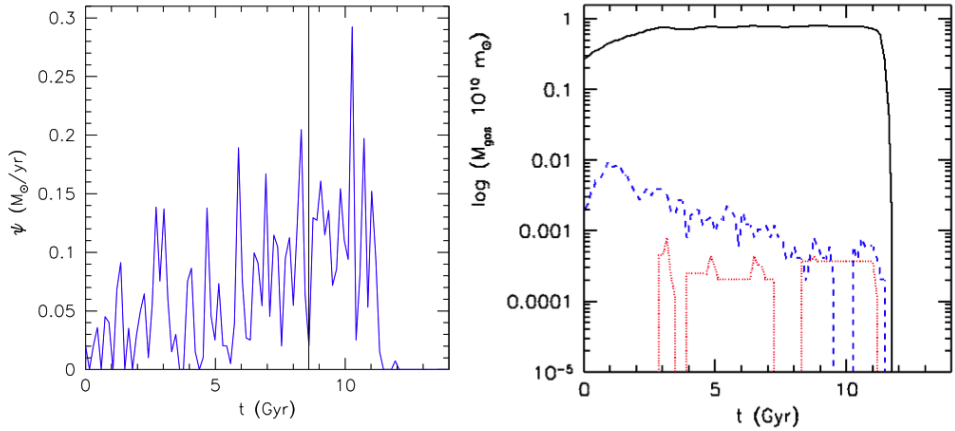

**Fig. 6.** Left**:** The evolution of the SFR predicted by our simulation for UGC 7639. The black vertical line corresponds to the galaxy age (8.6 Gyr; see text). Right: Gas accretion history (solid line), i.e., the evolution of the gas mass inside a radius of 50 kpc on the r-band image luminous center of the galaxy from our selected SPH simulations. The (blue) dashed and the (red) dotted lines correspond to the gas with temperature $10^4$ K and $10^6$ K, respectively.

in good agreement with age estimates from synthetic CMD comparisons of Tolstoy *et al.* (2009) for DGs in LV. This corresponds to a total B-band absolute magnitude of -15.6 mag, in good agreement with the data (Sect. 1 and Table 1). At this time our target lies about 0.5 Mpc away from the primary system, and the two galaxies are connected by a thin bridge of cold gas. We point out that this relative distance is greater than any projected distance on the sky plane. Karachentsev *et al.* (2013) report NGC 4258 as the main disturber, *i.e.*, the neighboring galaxy producing the maximal tidal influence on UGC 7639. In contrast, from our analysis, the Canes Venatici member whose total B-band magnitude agrees at best with our predictions turns out to be NGC 4242, with $M_B \cong -19$ mag (HyperLeda, *Makarov et al., 2014*), at a projected distance of 0.4 Mpc.

Figure 4 compares, on the same spatial scale and photometric band, the morphologies of our luminosity density map at the best-fit snapshot with the observed one. Figure 5, right panel, compares the r-band luminosity profile measured in Sect. 2.1 (blue circles) with that extracted from our simulated r-band luminosity density map (solid line) in Fig. 4 (left panel). The model profile has been derived with the same method (*i.e.*, using IRAF ellipse routine) used to obtain the observed profile (see Sec 2.1) and the same spatial resolution.

Figure 5, left panel, shows the comparison between the observed SED and our predictions. The amount of cold gas (*i.e.* T<20,000 K) expected inside the simulated map in Fig. 4 is about $8\times10^7$ $M_\odot$. Since the cooling time of this gas is much shorter than the snapshot time range (0.151 Gyr, Section 3), this amount represents the upper limit of the mass of cold gas, thus in agreement with the value derived from the Tully catalog (see Section 2.2).

The predicted FIR SED is composed by a warm and a cold dust component, both including polycyclic aromatic molecules, as described in Mazzei *et al.* (1992, 1994). Warm dust is located in a region of high radiation, *i.e.* in the neighborhood of OB stars (HII regions), whereas cold dust is heated by the interstellar diffuse radiation field. The distribution of diffuse radiation field best-fitting the FIR-SED is the same as described by Mazzei *et al.* (1992), *i.e.* a disk of gas and stars, in agreement with our findings in Sect. 2.1. The intensity field of such component is four times higher than the average in our own Galaxy, with a warm/cold energy ratio of 0.3. It follows that the contribution of the warm component to the dust luminosity, $L_{FIR}$ of UGC 7639 is about 25%, like that of our own Galaxy (*Mazzei et al. 1992*). A more detailed FIR coverage is required to derive strong conclusions concerning these points, however the greater intensity of the diffuse radiation field we derive agrees with the lower metallicity of stars in this dwarf compared with that in our galaxy. It is well known that low metallicity stars are more luminous than stars with solar composition (see *Chiosi 2007*, for a review). The average star metallicity provided by the best-fit snapshot for this DG is indeed 10 -20 times lower than that in the solar neighbour, *i.e.*, about 0.001. There are no measures of star metallicty available for this dwarf, however DGs are known as low-metallicity galaxies (*e.g. Tremonti et al., 2004*).

The current SFR at the selected snapshot is $\simeq 0.03\ M_\odot\ yr^{-1}$, in good agreement with UV-based estimates (see Section 2.2) accounting for a 2.4 factor due to the lower mass limit of the IMF in our simulation. The average age of galaxy population within $R_{25} \simeq 2$ kpc, weighted by B-band luminosity, is almost 1 Gyr, and the total mass inside the same radius is $5.3 \times 10^6\ M_\odot$, with a ratio between DM and gas+stars mass of about 10. The total mass of stars in Figure 4 is $1.9 \times 10^8\ M_\odot$ and the B-band M/L ratio results to be $4\ M_\odot / L_\odot$.

### 3.1.2. Evolution

Figure 6 (left panel) shows the behavior of the SFR that drives the evolution of UGC 7639. We want to highlight that the age of the simulation corresponding to our best-fit snapshot is 12.5 Gyr, since in the $10^{11}\ M_\odot$ halo of UGC 7639 the star formation starts about ≈4 Gyr after the beginning of the simulation. After its onset, the SFR increases, on average, up to a galaxy age of 10.3 Gyr (Fig. 6) when it reaches its maximum value, 10 times higher than that at the best-fit snapshot (8.6 Gyr, see above). Pacifici *et al.* (2013) found that low-mass galaxies have, on average, a rising star formation history, in good agreement with our predictions. Moreover, it is well known (*e.g. Bauer, 2013*) that low-mass galaxies, especially dwarfs, are susceptible to episodic (bursty) star formation events. This behavior is also well reproduced by our simulation. Note that the burst duration estimate is affected by time resolution and burst definition. Time resolution is 0.151 Gyr for our best-fit simulation (Sect 3). However, 302 Myr before the best-fit snapshot the SFR is about 10 times higher than its present-day value (see Sect. 3.1.1). Moreover, according to its oscillating behavior of Fig. 6 (left panel), the time range from a maximum to a minimum value of the SFR is, on average, 604 Myr, in good agreement with the average estimate by McQuinn *et al.* (2010).

Looking at the latest stages of the evolution, *i.e.* the remaining 4 Gyr before stopping our run, we noted that the burst is quenched after the gas is both exhausted and gradually removed (Fig. 6, right panel), while our dwarf is approaching more and more to the massive companion.

The behavior of the SFR cannot be provided by a simulation of the same halo evolving in isolation, as showed by Fig. 1 of MC03. In that case, the SFR neither increases on average, neither fades at the latest stage of the evolution.

The fate of our DG is thus to resemble that of a dwarf elliptical galaxy, devoid of gas and with red colors given the lack of young stars (Fig, 6). The path predicted in the rest-frame UV color-magnitude diagram (CMD) also highlights this fate (Fig. 7). The galaxy lies in a region corresponding to the extension of Wyder *et al.* (2007) blue sequence (BS, cyan line in Fig. 7) toward lower r-band luminosities. The BS is traced by the local spiral and starburst galaxies. By comparing Fig. 7 with Fig. 6, left panel, we note that our DG reaches its brightest r-band luminosity after 10.3 Gyr, *i.e.* when its SFR gets its maximum value. Fig. 6, right panel, shows that the SFR fades as the gas fueling decreases. The active phase of our dwarf lasts about 11 Gyr. Then, the galaxy crosses the green valley and reaches the red sequence (magenta line in Fig. 7), the locus of early-type galaxies, in two Gyrs, where it remains until the end of our simulation.

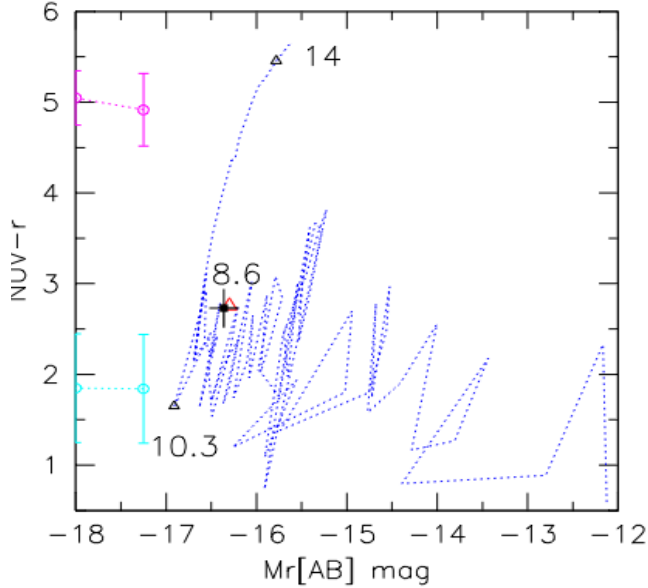

**Fig. 7.** The rest-frame evolution of ($NUV$ - $r$) vs M$_r$ color-magnitude diagram from our simulation of UGC 7639. Dotted line shows its evolutionary path; open symbols mark some meaningful evolutionary times in Gyr (see text); (red) triangle shows its current position. The observed position is given by the filled square, accounting for the same distance modulus as in Fig. 4, and the corresponding 1$\sigma$ uncertainty. The red (magenta) and the blue (cyan) sequences are also plotted following the same prescriptions as in Figure 3.

### 4. Summary & Conclusions

We used archive multi-wavelength data together with SPH simulations with chemo- photometric implementation to get insight into the present and evolutionary properties of the late-type dwarf UGC 7639 galaxy, a nearby DG hosting an inner star forming stellar population. We also investigated the UV-optical CMD of 22 galaxies in the Canes Venatici Cloud I, to which our target belongs, finding that UGC 7639 is among the faintest members, with the lowest cold gas amount.

We find that its global properties, namely its total absolute B-band magnitude, whole SED,

morphology, cold gas amount, and SFR, are well matched by an encounter with an object four times more massive. According to our data, the Canes Venatici member whose total B-band magnitude, $M_B \simeq -19$ mag agrees at best with our predictions turns out to be NGC 4242, which lies at a projected distance of 0.4 Mpc, in agreement with the simulation predictions.

The age of UGC 7639 we derived is 8.6 Gyr, whereas the average age of its stellar populations within $R_{25} \simeq 2$ kpc, weighted by B-band luminosity, is $\approx 1$ Gyr. Moreover, according to our simulation, its star formation will extinguish within 1.6 Gyr, thus likely leaving a dwarf elliptical galaxy.

As such, UGC 7639 is an example of DG whose global properties are well matched by our multi-wavelength and multi-technique approach, that seems to be very promising to explore the evolution of this class of galaxies (see the papers by Mazzei *et al.* 2014a,b, devoted to analyze evolution of early-type galaxies). We will exploit our approach further in future papers to highlight, in a fully consistent way, both the evolution of this couple of interacting galaxies and that of other DGs.

*Acknowledgements.* Funding for the SDSS and SDSS-II has been provided by the Alfred P. Sloan Foundation, the Participating Institutions, the National Science Foundation, the U.S. Department of Energy, the National Aeronautics and Space Administration, the Japanese Monbukagakusho, the Max Planck Society, and the Higher Education Funding Council for England. The SDSS Web Site is http://www.sdss.org/.
The Astrophysical Research Consortium manages the SDSS for the Participating Institutions. The Participating Institutions are the American Museum of Natural History, Astrophysical Institute Potsdam, University of Basel, University of Cambridge, Case Western Reserve University, University of Chicago, Drexel University, Fermilab, the Institute for Advanced Study, the Japan Participation Group, Johns Hopkins University, the Joint Institute for Nuclear Astrophysics, the Kavli Institute for Particle Astrophysics and Cosmology, the Korean Scientist Group, the Chinese Academy of Sciences ( LAMOST), Los Alamos National Laboratory, the Max-Planck-Institute for Astronomy (MPIA), the Max-Planck-Institute for Astrophysics (MPA), New Mexico State University, Ohio State University, University of Pittsburgh, University of Portsmouth, Princeton University, the United States Naval Observatory, and the University of Washington. This research has made use of the HyperLeda (http://leda.univ-lyon1.fr; Paturel et al. 2003) and the NASA/IPAC Extragalactic Database (NED), which is operated by the Jet Propulsion Laboratory, California Institute of Technology, under contract with the National Aeronautics and Space Administration.